# Stress-induced traps in multilayered structures

Magdalena Lidia Ciurea,[a] Sorina Lazanu, Ionel Stavarache, and Ana-Maria Lepadatu

National Institute of Materials Physics, 105 bis Atomistilor Street, Magurele 077125, Romania

Vladimir Iancu and Mihai Razvan Mitroi

"Politehnica" University of Bucharest, 313 Splaiul Independentei, Bucharest 060042, Romania

Raoul Rashid Nigmatullin

Kazan State University, Kazan 420008, Tatarstan, RUSSIA

Cristina Mihaela Baleanu

Mihail Sadoveanu Theoretical High School, Bucharest 021586, ROMANIA

The trap parameters of defects in $Si/CaF_2$ multilayered structures were determined from the analysis of optical charging spectroscopy measurements. Two kinds of maxima were observed. Some of them were rather broad, corresponding to "normal" traps, while the others, very sharp, were attributed to stress-induced traps. A procedure of optimal linear smoothing the noisy experimental data has been developed and applied. This procedure is based on finding the minimal value of the relative error with respect to the value of the smoothing window. In order to obtain a better accuracy for the description of the trapping-detrapping process, a Gaussian temperature dependence of the capture cross-sections characterizing the stress-induced traps was introduced. Both the normal and the stress-induced traps have been characterized, including some previously considered as only noise features.

---

[a] Electronic mail: ciurea@infim.ro

# 1. INTRODUCTION

The multilayered structures have many applications in micro-, nano- and optoelectronics. One of the problems that arise in these applications is the appearance of different kinds of defects at the interfaces. These defects act like traps and influence the electrical transport and phototransport in these structures.[1,2,3,4,5,6,7,8,9,10,11,12,13,14,15] For instance, the dark current is reduced as the carrier concentrations decrease due to the trapping. This effect is compensated in the case of the phototransport by the decrease of the recombination rate if either the electrons or holes are trapped. The lattice misfits and/or different dilatation coefficients may also induce local stresses. The local deformations produced by these stresses act as traps, too.[16,17,18,19,20] The interfacial traps can have specific applications, like the charge storage in non-volatile memories[21] and the improvement of the CMOS devices characteristics,[22] but they can also reduce the device reliability.[23,24,25]

Different methods were used to investigate these traps. The deep level transient spectroscopy (DLTS) method,[4,5,9,10,13,19,26] is the most used for the investigation of the deep traps. The method is based on the measurement of a transient junction capacitance and allows determining the energy of the trapping levels, as well as the concentration and capture cross-section of the trapping centers. One can also measure the transient conductance (conductance deep level transient spectroscopy – CDLTS).[27] An original alternative is the deep level photothermal spectroscopy (DLPTS) method,[28] which uses a super-band-gap optical excitation and a sub-band-gap optical monitoring of the non-equilibrium detrapped carriers. However, all these variants can be used only for non-homogeneous systems that present a space charge region.[29]

Another used method is the thermally stimulated currents (TSC) one.[1,2,30,31,32] In this method, the sample is cooled down, the traps are charged by an applied bias, an illumination, or both, and then the sample is heated up at low rate under a constant bias and the discharge current is measured. If the discharge is made under a non-zero bias, the relevant signal is represented by the difference between the discharge current after charging the traps and the zero current (no trap charging) measured under the same applied bias. If the discharge is made under zero bias, the relevant signal is the discharge current, also named relaxation current (RC). From the modeling of the discharge process, the trap parameters can be obtained. The most interesting variant of this method is the optical charging spectroscopy (OCS),[31,32] where both the (optical) charging and the (thermal) discharging of the traps are made under zero bias.

One of the problems that arise in analyzing the experimental data is that they are rather noisy.[26,28] Consequently, both the general shape of the experimental curves and the maxima identification (position, amplitude and width) become difficult. A procedure for the noise smoothing that will not eliminate significant maxima is then necessary to properly interpret the experimental results. Most of the smoothing procedures discussed in the literature[33,34,35,36,37,38,39,40,41] present the following drawbacks: (1) the absence of a general criterion that helps to select the most suitable smoothing procedure among others; (2) some procedures cannot be applied if the initial data have outliers; (3) some procedures use nonlinear transformations; (4) many suggested procedures have uncontrollable errors that are difficult to evaluate; (5) the absence of a general criterion for the selection of the proper wavelet function[33,42] makes these wavelet transform methods incompatible with each other (when different types of wavelet functions are used).

The present paper analyzes the stress-induced traps in Si/CaF$_2$ multilayered structures, determined by OCS measurements. As the experimental data are noisy, the first step is to filter them. Then the smoothed discharge current is modeled and the trap parameters are deduced.

## 2. EXPERIMENTAL RESULTS

The multi-quantum wells (MQW) samples were made of $q = 50$ bilayers of Si/CaF$_2$, with equal thickness ($g_{Si} = g_{CaF_2} \equiv g = 1.6$ nm).[32,43,44] They were deposited on n-type (111) Si substrate, by using the molecular beam epitaxy. The samples were provided with Al electrodes – a bottom Ohmic one and a top semitransparent one. Due to this sandwich configuration, all the investigated quantities are averaged over the (horizontal) top electrode area $A$:[32]

$$Q(z) \equiv \frac{1}{A} \iint_A Q(x, y, z) dx dy. \tag{1}$$

The traps were studied by means of the OCS method.[31,32] In this zero-bias method, the sample is cooled down at a low temperature $T_0$. The stress that appears during the cooling, as a consequence of the different dilatation coefficients of the Si and CaF$_2$ layers induces a number of interface traps. Their parameters depend on the cooling regime. Other interface traps appear due to the local interactions. Then, the sample is illuminated with monochromatic light in the absorption band and the photogenerated carriers are trapped. These trapped carriers produce a frozen-in electric field. After the charging of the traps, the light is switched off, the sample is connected to an electrometer and it is heated up. The heating is made at a constant low rate β. The detrapped carriers move in the frozen-in field of the still trapped ones and the discharge current is measured as function of the temperature.

The measurements used $T_0$ = 77 K, β = 0.1 K/s, and two excitation wavelengths, 0.5 and 1.0 μm. The penetration depths of these wavelengths are 5 and 1000 μm, respectively.[45] Under such conditions, one can consider that almost all traps are filled up after the illumination. The obtained curves are presented in Fig. 1. One can see that the recorded signal is noisy, so that it is necessary to filter it before any information is extracted from the curves.

## 3. THE PROCEDURE OF THE OPTIMAL LINEAR SMOOTHING

For the smoothing of the initial data we applied the procedure of optimal linear smoothing (POLS).[46,47,48] This method helps to find the optimal and smoothed trend (the so-called pseudo-fitting function) and separate it from the relative fluctuations. Minimizing the value of the relative error on the plot "relative error with respect to the value of the smoothing window" in the vicinity of the first local minima, one can obtain the pseudo-fitting function without information related to the physical model used. Besides this important peculiarity the new method enables to read the remaining detrended noise and express the desired distribution in terms of the fitting parameters corresponding to the envelope of the sequence of the ranged amplitudes (SRA).[47]

Usually, any random data sequence considered, $y_j = y(x_j)$, contains both large-scale fluctuations (that represent the trend) and high-frequency fluctuations (that represent the noise). They can be separated from each other, by applying the procedure of the optimal linear smoothing (POLS). In this procedure, one defines the smoothed data sequence as:

$$\tilde{y}_j(w) \equiv \frac{\sum_{j'=1}^{N} K\left(\frac{x_j - x_{j'}}{w}\right) y_{j'}}{\sum_{j'=1}^{N} K\left(\frac{x_j - x_{j'}}{w}\right)}, \quad K(t) = \exp\left(-\frac{t^2}{2}\right), \quad (2)$$

where $K(t)$ defines the Gaussian kernel and $w$ is the initial value of the smoothing window, chosen large enough to start the procedure. The set $y_j$ ($j = 1, 2, …, N$) represents the initial noisy sequence, while the set $\tilde{y}_j(w)$ ($j = 1, 2, …, N$) is the sequence smoothed on the window $w$. The choice of a Gaussian kernel has two important advantages: (a) the smoothed function given by Eq. (2) is obtained as the result of a linear transformation that does not introduce uncontrollable errors; (b) the value of the smoothing window $w$ is an adjustable (fitting) parameter and may take any value. The function $\tilde{y}_j(w)$ is a pseudo-fitting function, which is not associated directly with a specific model describing the investigated process. The value of the initial fluctuations is decreased by means of an iterative procedure, reducing the smoothing window. The optimal smoothing window value, $w_{opt}$, is found by minimizing the value of the relative error, $RelErr(w')$, defined by the relations:

$$\Delta y_j(w') \equiv y_j - \tilde{y}_j(w'), \quad w' < w,$$

$$StDev(\Delta y(w')) \equiv \sqrt{\frac{1}{N-1} \sum_{j=1}^{N} \Delta y_j^2(w')},$$

$$Mean(\tilde{y}(w')) \equiv \frac{1}{N} \sum_{j=1}^{N} \tilde{y}_j(w'), \tag{3}$$

$$RelErr(w') \equiv \left[ \frac{StDev(\Delta y(w'))}{Mean(\tilde{y}(w'))} \right] \cdot 100\,\%,$$

in the vicinity of the first local minimum.

It is necessary to emphasize the following fact. The behavior of the function $RelErr(w')$, investigated for many different random sequences with hidden (uncertain) trend has some specific features. This function has at least three minimal points. The first minimal point coincides with the global minimum when the value of $w$ is very close to the given discrete step ($h = \Delta x$). The third minimal point coincides with the large values of $w$, corresponding to the mean value of the random sequence analyzed. The first local minimum can appear after the global minimum. The desired trend that minimizes the value of the relative error is given by the expression:

$$\tilde{y}_j \equiv \tilde{y}_j(\tilde{w}), \quad \tilde{w} \equiv w_{opt}. \tag{4}$$

After finding the optimal smoothing window value, it becomes possible to separate the initial random sequence in two parts: (a) the optimal trend, given by Eq. (4), and (b) the detrended sequence of the relative fluctuations:

$$Srf_j = \frac{\Delta y_j(\tilde{w})}{\tilde{y}_j}. \tag{5}$$

The application of the POLS on the OCS data is illustrated in Figs. 2 – 6. Real data with their smoothed trends are presented in Figs. 2 and 3. On these figures we show two original data files (defined as OCS discharge curves for two wavelengths: 0.5 and 1.0 μm). The optimal trends are shown by solid lines. Figure 4 shows the plot presented by expression (2). The values of the minimal error correspond to the first local minima and their optimal smoothing windows are shown by arrows. Figures 5 and 6 show the behavior of the relative fluctuations. The "spikes" on this figure are helpful for the detection of the peaks that can be hidden on the initial (noisy) data analyzed.

The final filtered discharge curves are presented in Fig. 7. The different maxima and shoulder on both curves are numbered function of the temperature. One can see that some of them (No. 1, 4, and 5) appear as typical trap discharge maxima, some (No. 2, 3, 6, 7, 8, and 9) are very sharp spikes, due to the stress, and the final tail (No. 10) is due to the capacitive behavior of the sample.

## 4. MODELING AND DISCUSSION

The detrapping process was discussed in detail in a previous paper.[32] The equations that govern this process are:

$$\left\{1+\sum_{i'}\tau_{ni'}c_{ni'}(T)[N_{ti'}(z,T)-n_{ti'}(z,T)]\right\}\frac{\partial}{\partial T}n_{ti}(z,T)=-\frac{1}{\beta}c_{ni}(T)N_{cti}(T)n_{ti}(z,T),$$
$$\left\{1+\sum_{j'}\tau_{pj'}c_{pj'}(T)[P_{tj'}(z,T)-p_{tj'}(z,T)]\right\}\frac{\partial}{\partial T}p_{tj}(z,T)=-\frac{1}{\beta}c_{pj}(T)P_{vtj}(T)p_{tj}(z,T),$$
(6)

$$\Delta n(z,T) \equiv \sum_i \Delta n_i(z,T) = \sum_i \tau_{ni} c_{ni}(T) N_{cti}(T) n_{ti}(z,T),$$
$$\Delta p(z,T) \equiv \sum_j \Delta p_j(z,T) = \sum_j \tau_{pj} c_{pj}(T) P_{vtji}(T) p_{tj}(z,T),$$
(7)

where the following notations were made:

- $n_{ti}(z,T)$, $p_{tj}(z,T)$ are the trapped carrier concentrations for the levels $i$ and $j$, with trap concentrations $N_{ti}(z,T)$ and $P_{tj}(z,T)$, respectively;

- $c_{ni}(T) = \sigma_{ni}(T)\tilde{v}_e(T)$, $c_{pj}(T) = \sigma_{pj}(T)\tilde{v}_h(T)$ are the capture coefficients, with $\sigma_{ni}(T)$, $\sigma_{pj}(T)$ as the capture cross-sections, and $\tilde{v}_e(T)$, $\tilde{v}_h(T)$ the mean electron and hole velocities;

- $N_{cti}(T) = 2\left(\dfrac{m_e^* k_B T}{2\pi\hbar^2}\right)\exp\left(-\dfrac{E_c - E_{ti}}{k_B T}\right)$, $P_{vtj}(T) = 2\left(\dfrac{m_h^* k_B T}{2\pi\hbar^2}\right)\exp\left(\dfrac{E_v - E_{tj}}{k_B T}\right)$, with $E_{ti}$, $E_{tj}$ the trap activation energies;

- $\Delta n = n - n_0$, $\Delta p = p - p_0$ are the (non-equilibrium) detrapped carrier concentrations ($n_0$ and $p_0$ are the equilibrium concentrations, respectively).

As the investigated samples are homogeneous, the $z$ dependence can be discarded. On the other hand, the trap and trapped carrier concentrations in Eqs. (6, 7) are defined as volume concentrations. The surface concentrations can be obtained by multiplying these values by the layer thickness $g$.

While the normal trap concentrations and capture cross-sections are practically independent on the temperature, the stress ones are in a different situation. As they are due to the differences between the dilatation coefficients, it is to be expected that their concentrations should depend on the temperature as:

$$N_{ti}(T) = N_{ti}^{(0)}\left(1 - \dfrac{T}{T_s}\right)^{\gamma_i}, \quad P_{tj}(T) = P_{tj}^{(0)}\left(1 - \dfrac{T}{T_s}\right)^{\gamma_j}, \tag{8}$$

where $T_s$ is the storage temperature. Concerning the capture cross-sections, they should be correlated with a specific temperature, corresponding to the stress trap discharge, i.e. the spike position. The simplest way to describe this temperature dependence is to consider the cross-sections as proportional with a probability distribution, centered on the spike position. As the thermal fluctuations are described by a Gaussian, the stress trap capture cross-sections were taken in the form:

$$\sigma_{ni} = \sigma_{ni}^{(0)} \exp\left(-\frac{(T-T_{ni})^2}{2W_{ni}^2}\right), \quad \sigma_{pj} = \sigma_{pj}^{(0)} \exp\left(-\frac{(T-T_{pj})^2}{2W_{pj}^2}\right). \qquad (9)$$

To reduce the mathematical intricacies, the final tail was treated as a normal trap.

The differential equations (6) form a non-linear coupled system. This system, together with the equations (7), was numerically integrated by means of an original Mathcad 14.0 program that uses the Rkadapt solver, based on the fourth-order Runge-Kutta algorithm, with adaptive step-size. The initial values of the activation energies, the capture cross sections and the concentrations of the traps were chosen near the values reported in Ref. 32 for the traps whose maxima are located at approximately the same temperatures. For the others, the initial values of the activation energies were chosen from the Bube formula ($E_{ti}/k_B T_{ti} \approx \text{const}$, where $T_{ti}$ is the temperature of the maximum discharge current for the considered trap), and the other parameters were taken of the same order of magnitude as those for normal traps.

After solving the system, the program computed the mean value of the frozen-in electric field as:

$$\tilde{E}(T) = \frac{2q+1}{4}\frac{eg}{\varepsilon_0\varepsilon_r}[p_t(T)-n_t(T)] \equiv \frac{2q+1}{4}\frac{eg}{\varepsilon_0\varepsilon_r}\left[\sum_j p_{tj}(T)-\sum_i n_{ti}(T)\right], \qquad (10)$$

where $\varepsilon_r$ is the CaF$_2$ relative permittivity (by definition, $e > 0$). Then the mean bias per CaF$_2$ layer is $\widetilde{U}(T) = 2g\widetilde{E}(T)$. These layers are insulators, so that only displacement and tunneling currents can pass through the structure. The currents are computed as:

$$I_d = \varepsilon_0 \varepsilon_r A \beta \frac{d\widetilde{E}}{dT}, \qquad (11)$$

$$I_t = \mathrm{sign}(\widetilde{U})\left\{ I_n\left[\left(1 - \frac{|\widetilde{U}|}{U_n}\right)\exp\left(-\alpha_n\sqrt{1 - \frac{|\widetilde{U}|}{U_n}}\right) - \exp(-\alpha_n)\right] \right.$$
$$\left. - I_p\left[\left(1 + \frac{|\widetilde{U}|}{U_p}\right)\exp\left(-\alpha_p\sqrt{1 + \frac{|\widetilde{U}|}{U_p}}\right) - \exp(-\alpha_p)\right] \right\}, \qquad (12)$$

with $I_n = enA\widetilde{v}_e/\sqrt{3}$, $I_p = epA\widetilde{v}_h/\sqrt{3}$, $\alpha_n = \sqrt{8em_e^* U_n g^2/\hbar^2}$, $\alpha_p = \sqrt{8em_h^* U_p g^2/\hbar^2}$, $eU_n$ and $eU_p$ being the heights of the tunneled barriers for electrons and holes. As the sample is homogeneous, the diffusion current is null, so that the total discharge current is $I = I_d + I_t$.

By adjusting the trap parameters, the discharge current is made to fit the experimental data. The results are presented in Figs. 8, 9 and Table I. The maxima positions $(T_n, T_p)$ are marked for the stress-induced spikes only. The $\gamma$ exponents in Eqs. (8) and the half-widths in Eqs. (9) are taken equal and independent on the trapped charge sign ($\gamma_n = \gamma_p = 4$, $W_n = W_p = \sqrt{2}$ K), as the primary cause for all the spikes is the same. The lifetimes for the all the detrapped electrons, and all the detrapped holes, respectively, are also taken equal ($\tau_n = 400$ ns, $\tau_p = 180$ ns). One can see that not all the trapping levels are common for the two curves and, for those which are common, the trap concentrations appear lower for $\lambda = 1.0$ μm (lower absorption and therefore lower filling).

## 5. CONCLUSIONS

Trapping phenomena in Si/CaF$_2$ multi-quantum well structures were investigated by the OCS method. The initially noisy data have been filtered using an original smoothing procedure, the POLS, which is general, linear and does not introduce uncontrollable errors. The application of this method leaded to the elimination of both large-scale and high-frequency fluctuations. The optimal smoothing of the initial data helped us to identify traps that were previously considered as noise features.

Beside normal traps, whose concentrations and capture cross-sections are independent on the temperature, we also considered the stress-induced ones. As they are due to the differences between the dilatation coefficients, their concentrations were considered as temperature dependent. Their capture cross-sections were correlated with the specific temperature, corresponding to the stress-induced trap discharge, i.e. the spike position. The correlation was described by introducing a probability distribution, centered on the spike position, as a proportionality factor. The factor was chosen to be Gaussian, in agreement with the thermal fluctuations behavior. This way, the experimental data were well fitted by the model, and the characteristics of both normal and stress-induced traps were extracted.

## ACKNOWLEDGEMENTS

The Romanian contribution to this work was supported by CNCSIS –UEFISCSU, project number PNII – IDEI 918/2008. The Russian contribution was supported by the Russian

Ministry of Higher Education from the grant Russian Scientific Potential of Higher Schools – RSP-14.14

**Table I. Trap parameters resulted from the fitting procedure.**

| Trap No. | Trap type | $\lambda$ (μm) | $\sigma^{(0)}$ ($10^{-18}$ cm$^2$) | $T_n, T_p$ (K) | $N_t^{(0)}, P_t^{(0)}$ ($10^{15}$ cm$^{-3}$) | $E_t$ (eV) |
|---|---|---|---|---|---|---|
| 1 | n | 0.5 | 1.3 | – | 0.375 | 0.30 |
| 1 | n | 1.0 | 1.4 | – | 0.15 | 0.30 |
| 2 | n | 0.5 | 50 | 192 | 3.2 | 0.36 |
| 3 | n | 0.5 | 50 | 202 | 1.3 | 0.38 |
| 4 | n | 1.0 | 1.0 | – | 0.28 | 0.40 |
| 5 | n | 0.5 | 1.0 | – | 0.08 | 0.41 |
| 6 | n | 0.5 | 50 | 229 | 12.5 | 0.42 |
| 6 | n | 1.0 | 50 | 229 | 6.8 | 0.42 |
| 7 | p | 0.5 | 50 | 232 | 4.1 | 0.44 |
| 7 | p | 1.0 | 50 | 232 | 0.9 | 0.44 |
| 8 | n | 1.0 | 50 | 236 | 3.14 | 0.45 |
| 9 | p | 0.5 | 50 | 239 | 3.5 | 0.45 |
| 10 | n | 0.5 | 1.6 | – | 75 | 0.63 |
| 10 | n | 1.0 | 1.6 | – | 8.5 | 0.63 |

# Figure Captions

**Fig. 1 (color online). OCS discharge current (as measured) for excitation wavelength $\lambda = 0.5$ μm (open circles) and $\lambda = 1$ μm (open stars).**

**Fig.2 (color online). Application of the POLS for OCS file, $\lambda = 0.5$ μm. The optimal trend (SmOCS) is shown by solid line.**

**Fig.3 (color online). Application of the POLS for OCS file, $\lambda = 1.0$ μm. The optimal trend (SmOCS) is shown by solid line.**

**Fig. 4 (color online). The behavior of the relative error with respect to the value of the current window. The optimal values of the window together with minimal values of the relative errors for $\lambda = 0.5$ μm (open circles) and $\lambda = 1.0$ μm (open stars) are shown by arrows.**

**Fig. 5. The relative fluctuations obtained for $\lambda = 0.5$ μm. The spikes shown on the detrended noise help to determine the localization of the peaks hidden on original data.**

Fig. 6. The relative fluctuations obtained for λ = 1.0 μm. The spikes shown on the detrended noise help to determine the localization of the peaks hidden on original data.

Fig. 7 (color online). Filtered OCS discharge current for excitation with λ = 0.5 μm (open circles) and λ = 1 μm (open stars).

Fig. 8 (color online). Fitting results for λ = 0.5 μm: open stars – filtered curve; solid line – model fit.

Fig. 9 (color online). Fitting results for λ = 1 μm: open stars – filtered curve; solid line – model fit.

[26] E. Płaczek-Popko, J. Trzmieł, E. Ziełony, S. Grzanka, R. Czernecki, and T. Suski, Physica B **404**, 4889 (2009).

[27] M. Gassoumi, J. M. Bluet, F. Chekir, I. Dermoul, H. Maaref, G. Guillot, A. Minko, V. Hoel, and C. Gaquière, Mater. Sci. Eng. C **26**, 383 (2006).

[28] A. Mandelis and J. Xia, J. Appl. Phys. **103**, 043704 (2008).

[29] D. Cavalcoli, A. Cavallini, M. Rossi, and S. Pizzini, Semiconductors **41**, 421 (2007).

[30] O. V. Brodovoy, V. A. Skryshevsky, and V. A. Brodovoy, Solid-State Electron. **46**, 83 (2002).

[31] V. Iancu, M. L. Ciurea, and M. Draghici, J. Appl. Phys. **94**, 216 (2003).

[32] M. L. Ciurea, V. Iancu, and M. R. Mitroi, Solid-State Electron. **51**, 1328 (2007).

[33] V. J. Barclay and R. F. Bonner, Anal. Chem. **69**, 78 (1997).

[34] Z.-Y. Cai and M.-Z. Li, Measurement **33**, 47 (2002).

[35] M. Laddomada, Signal Proc. IET **4**, 158 (2010).

[36] J. Zhang, H. F. Chen, F. Fang, and W. Liao, IEEE Trans. Biomed. Eng. **57**, 343 (2010).

[37] M. Jakubowska, E. Hull, R. Piech, and W. W. Kubiak, Chem. Anal. **53**, 215 (2008).

[38] K.-J. He and Z.-M. Chen, Comput. Appl. **27**, 1479 (2007).

[39] J. Wang, L. S. Zhou, and L. Y. Zhang, China Mech. Eng. **17**, 1744 (2006).

[40] C. H. Chou, Physica B - Cond. Matter **253**, 320 (1998).

[41] R. Ruotolo and D. M. Storer, J. Sound Vibr. **239**, 41 (2001).

[42] C. E. Heil and D. F. Walnut, SIAM Rev. **31**, 628 (1989).

[43] V. Ioannou-Sougleridis, V. Tsakiri, A. G. Nassiopoulou, F. Bassani, S. Menard, and F. Arnaud d'Avitaya, Mater. Sci. Eng. B **69-70**, 309 (2000).
20

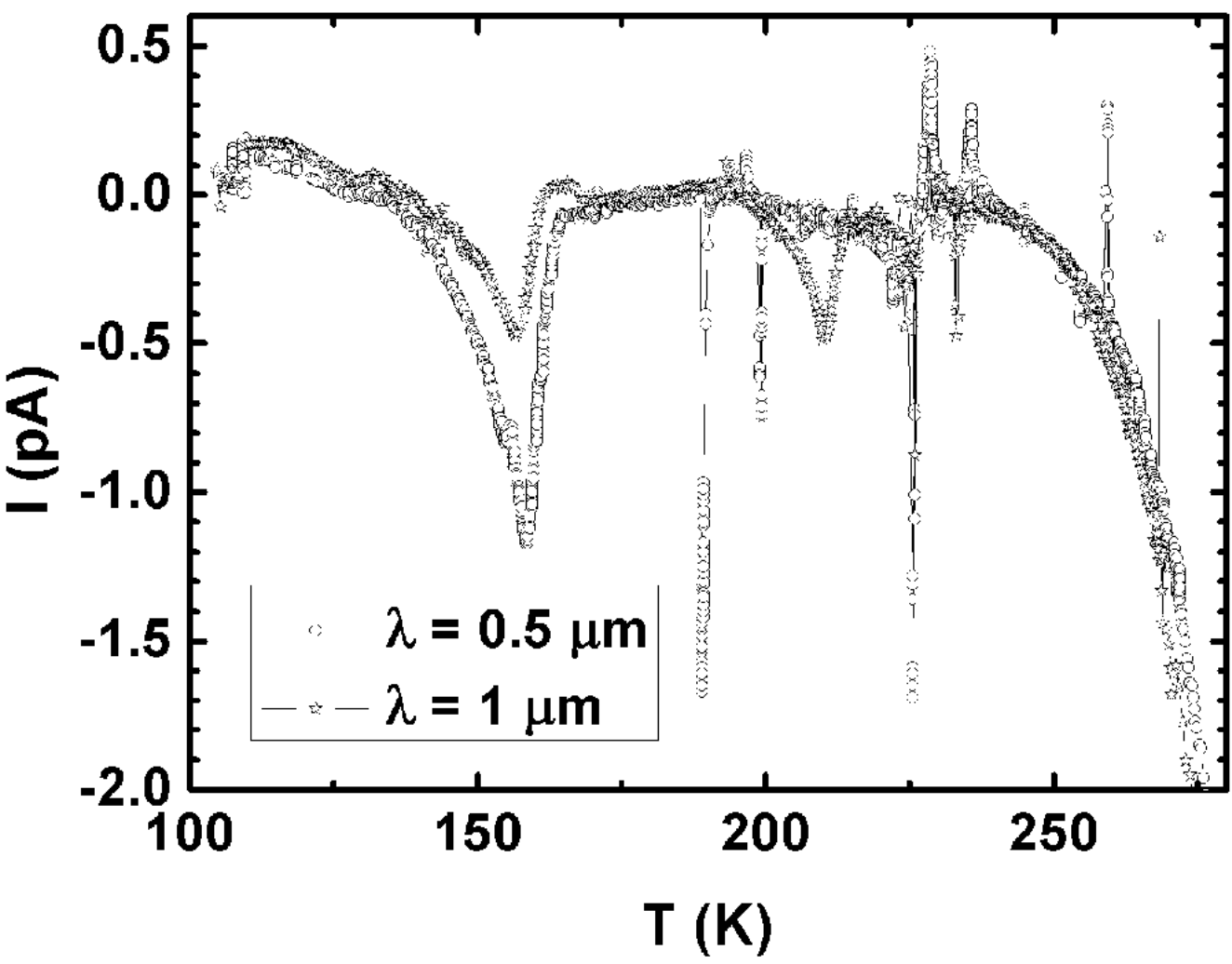

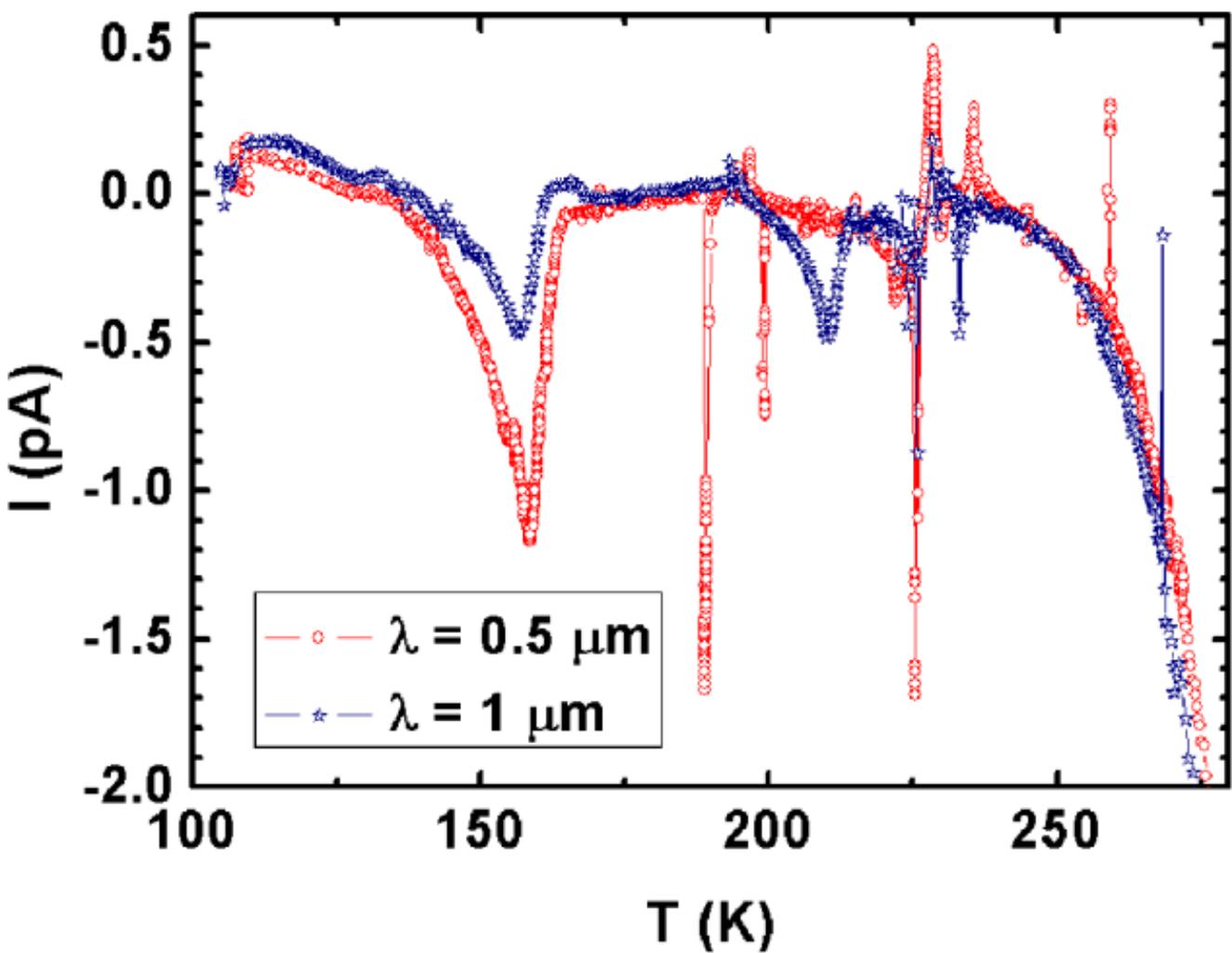

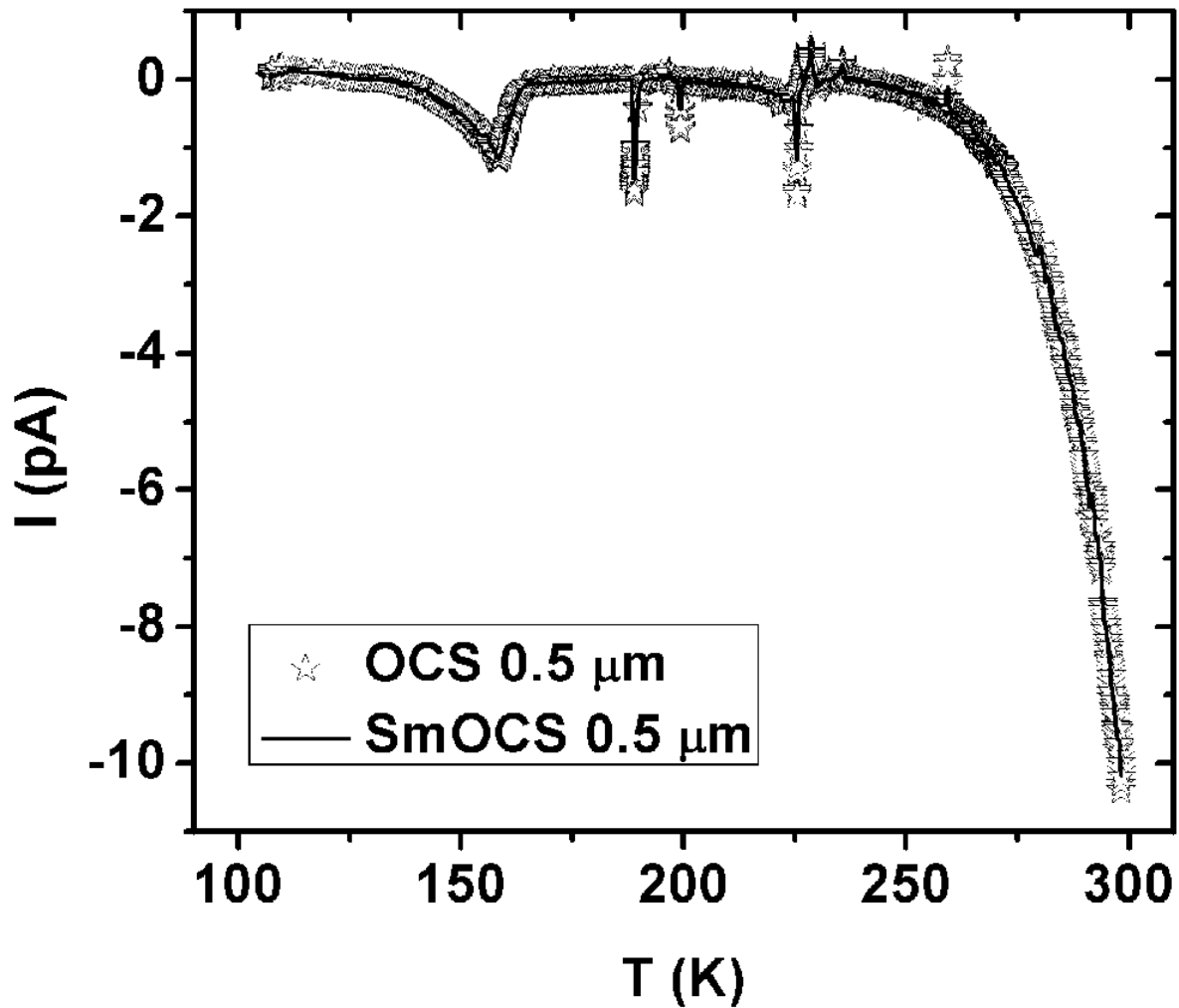

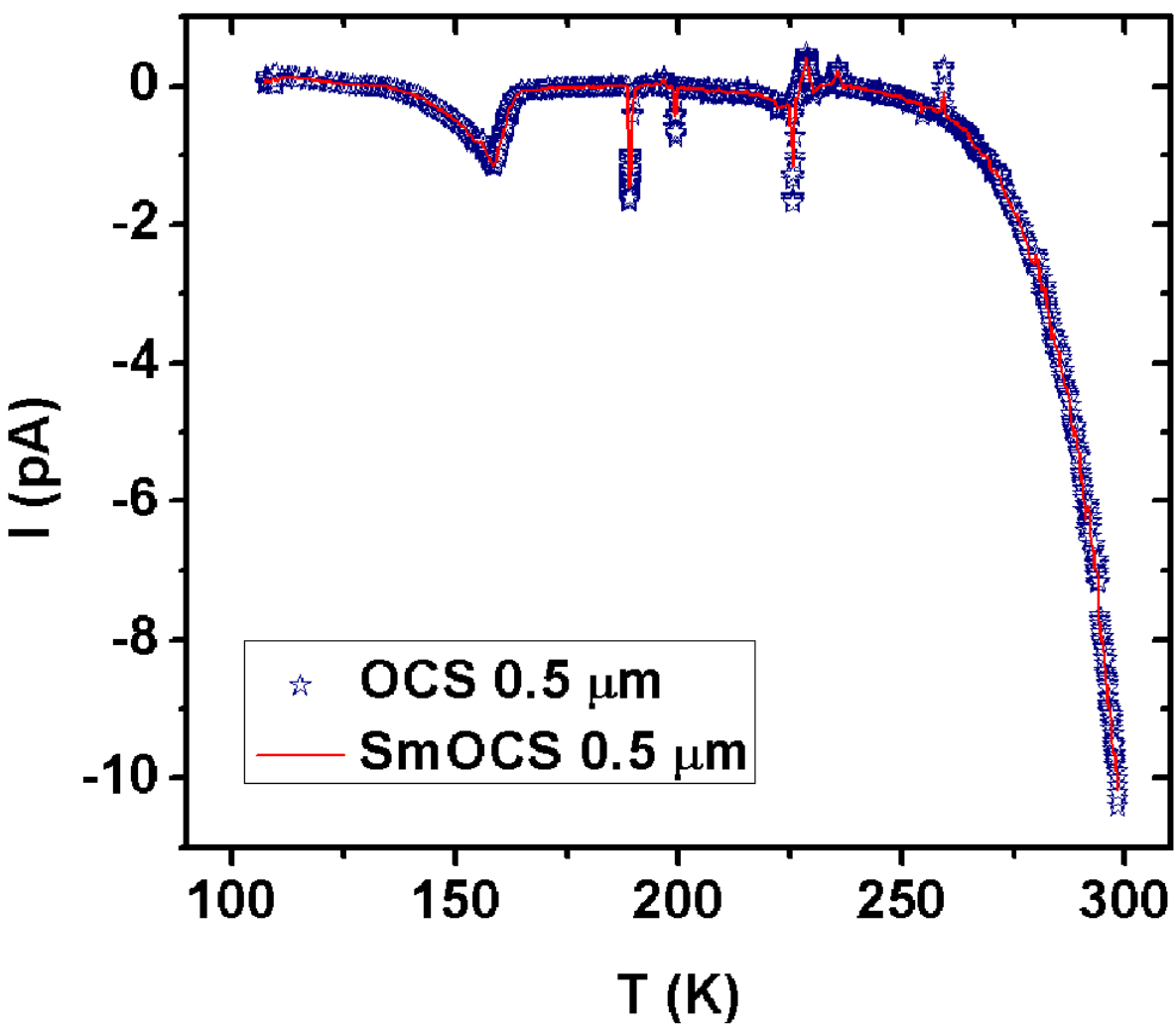

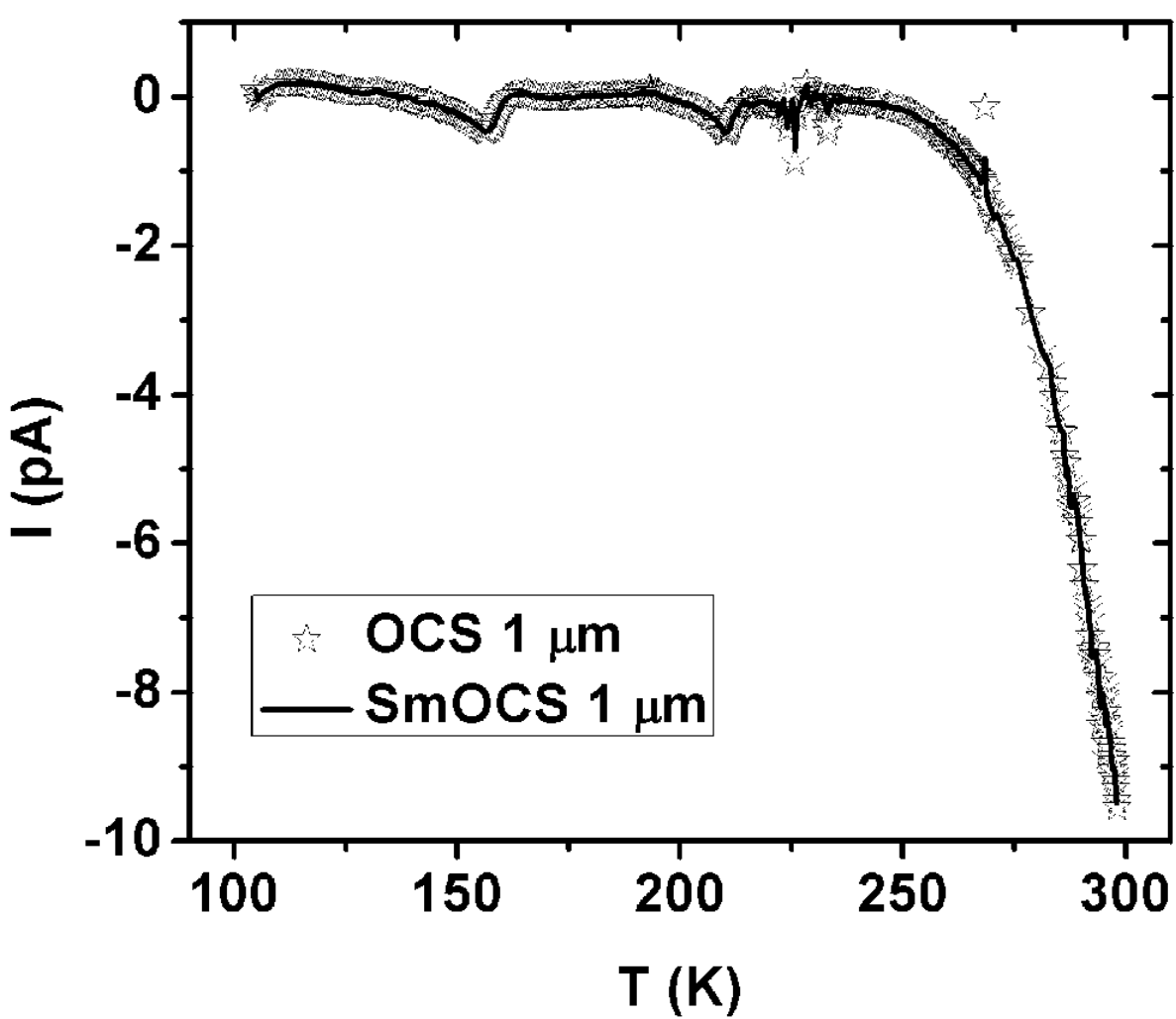

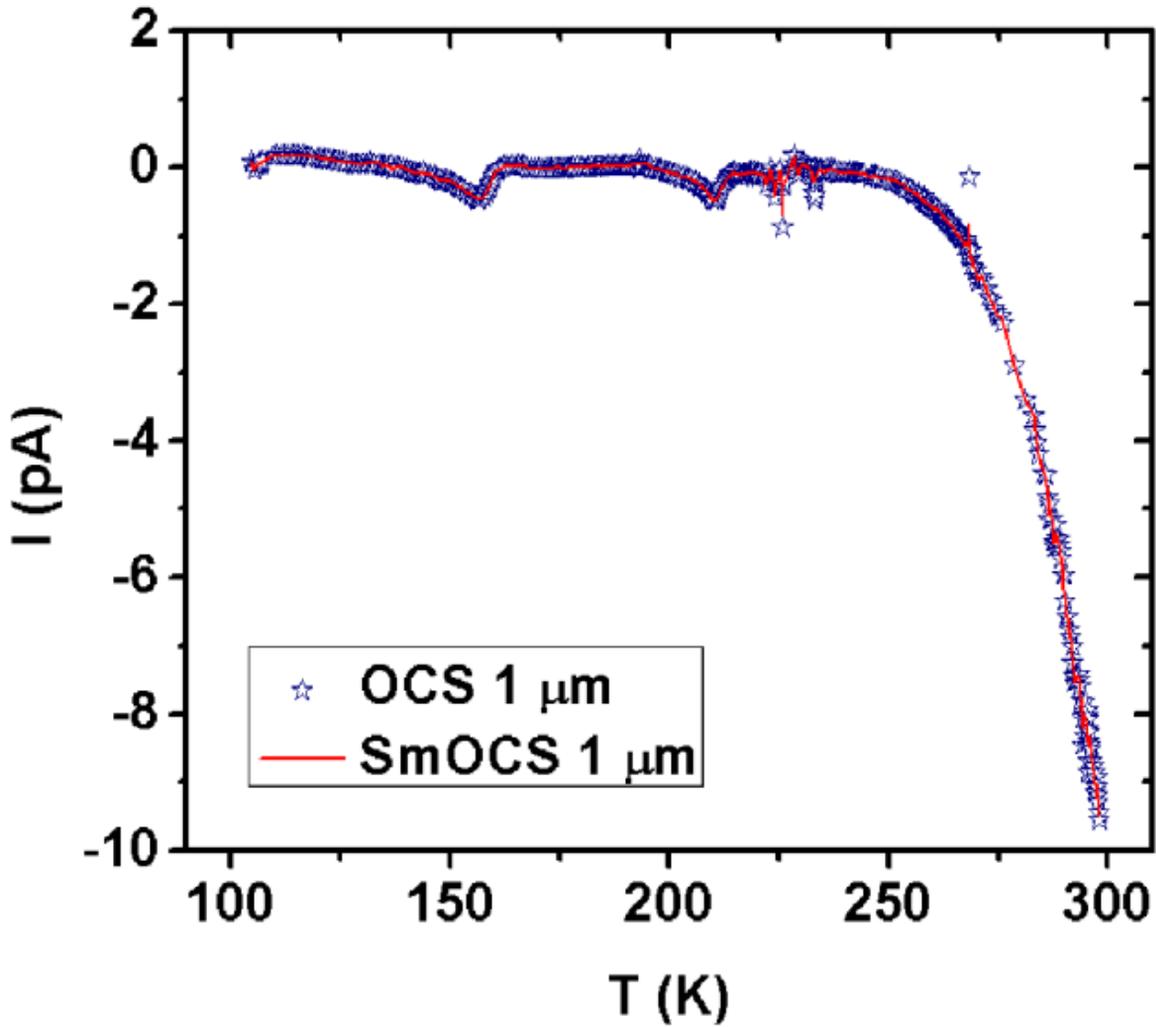

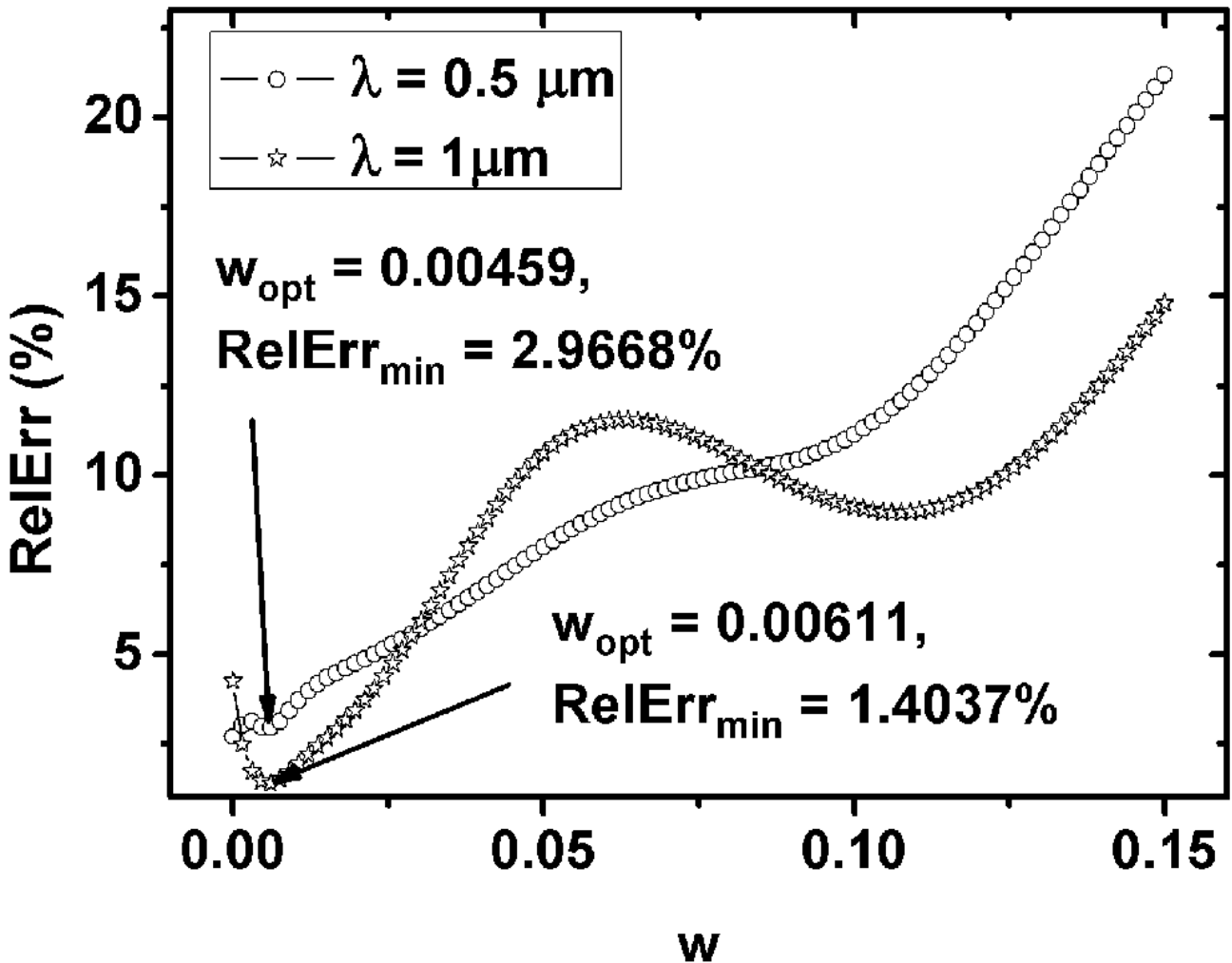

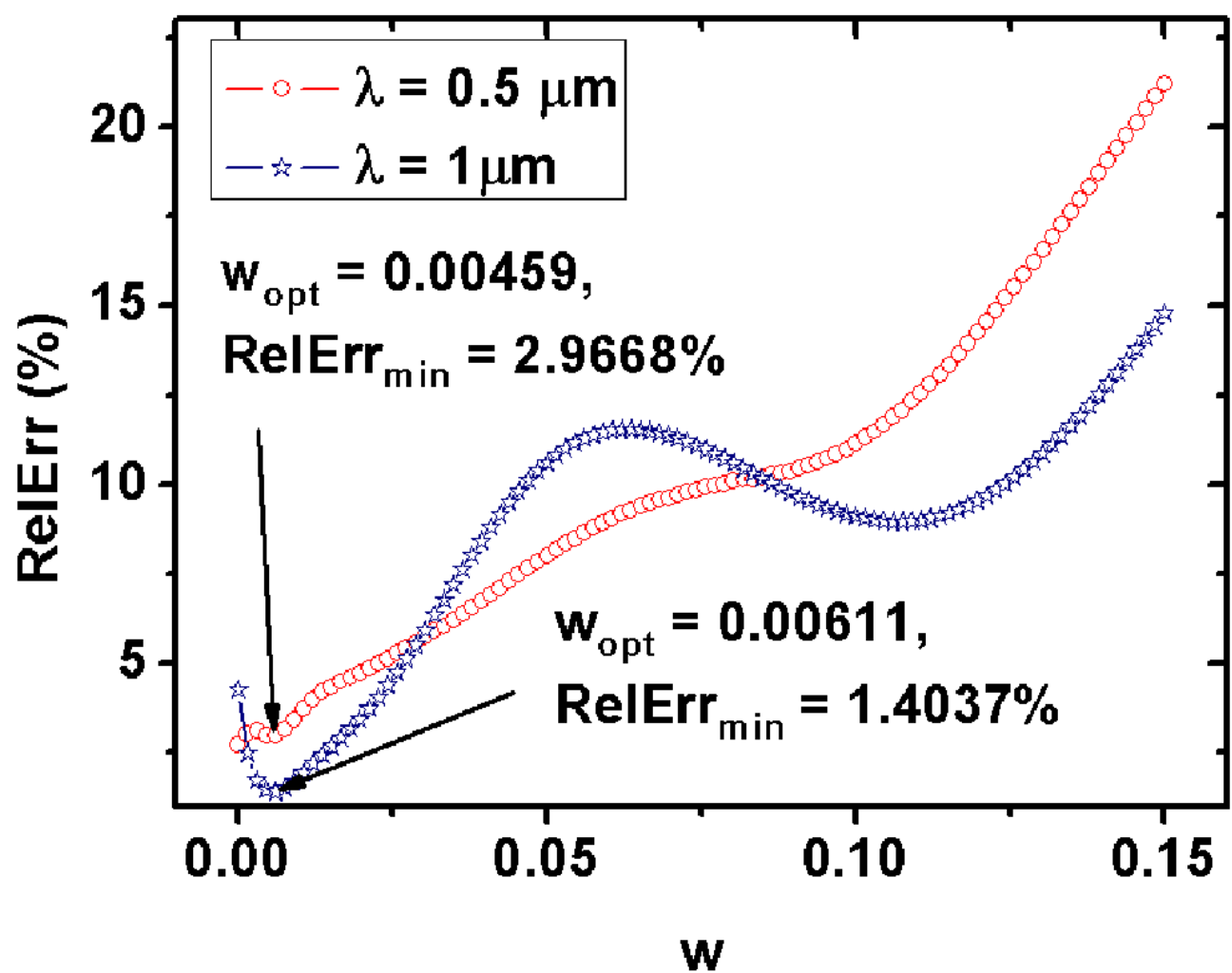

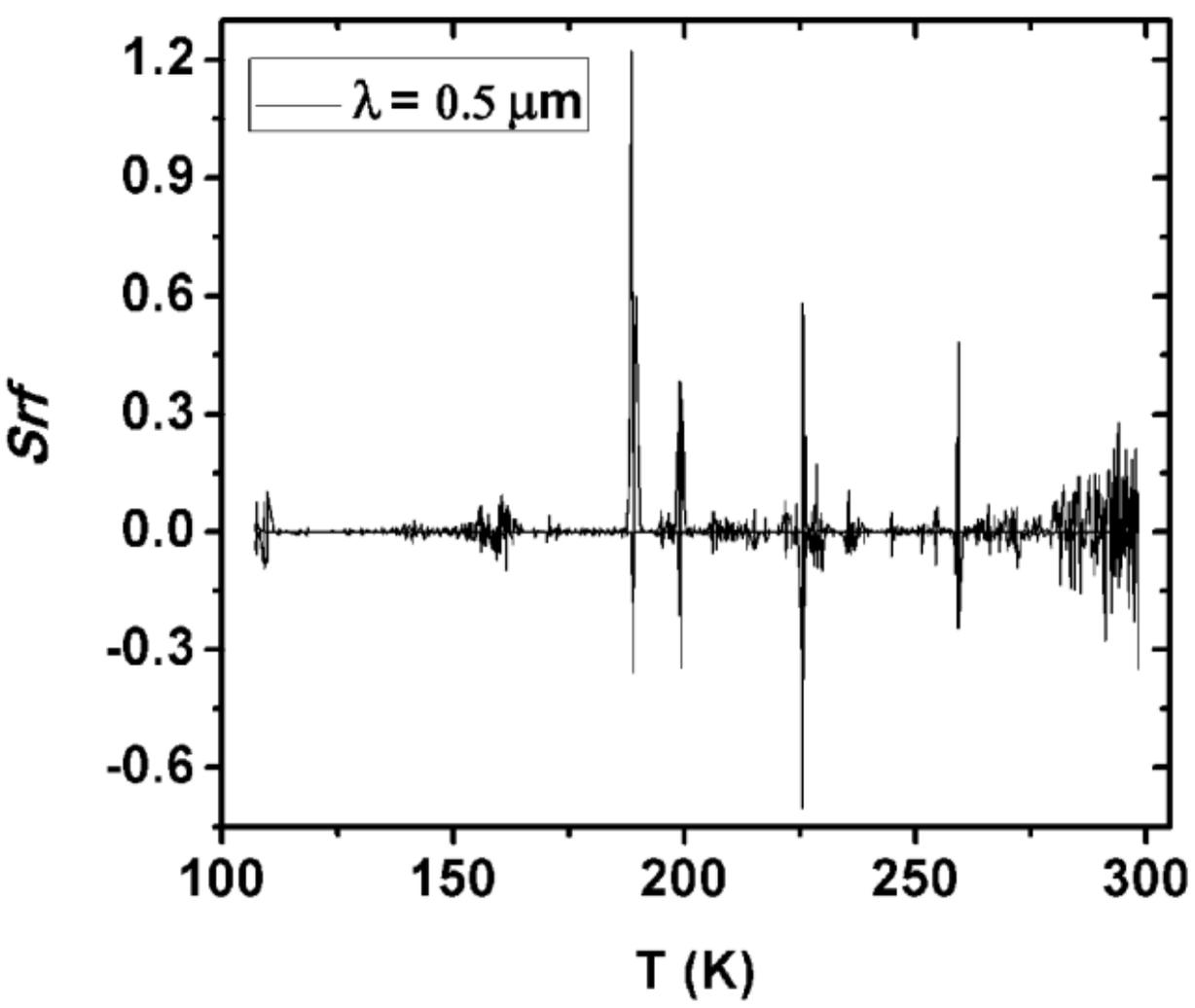

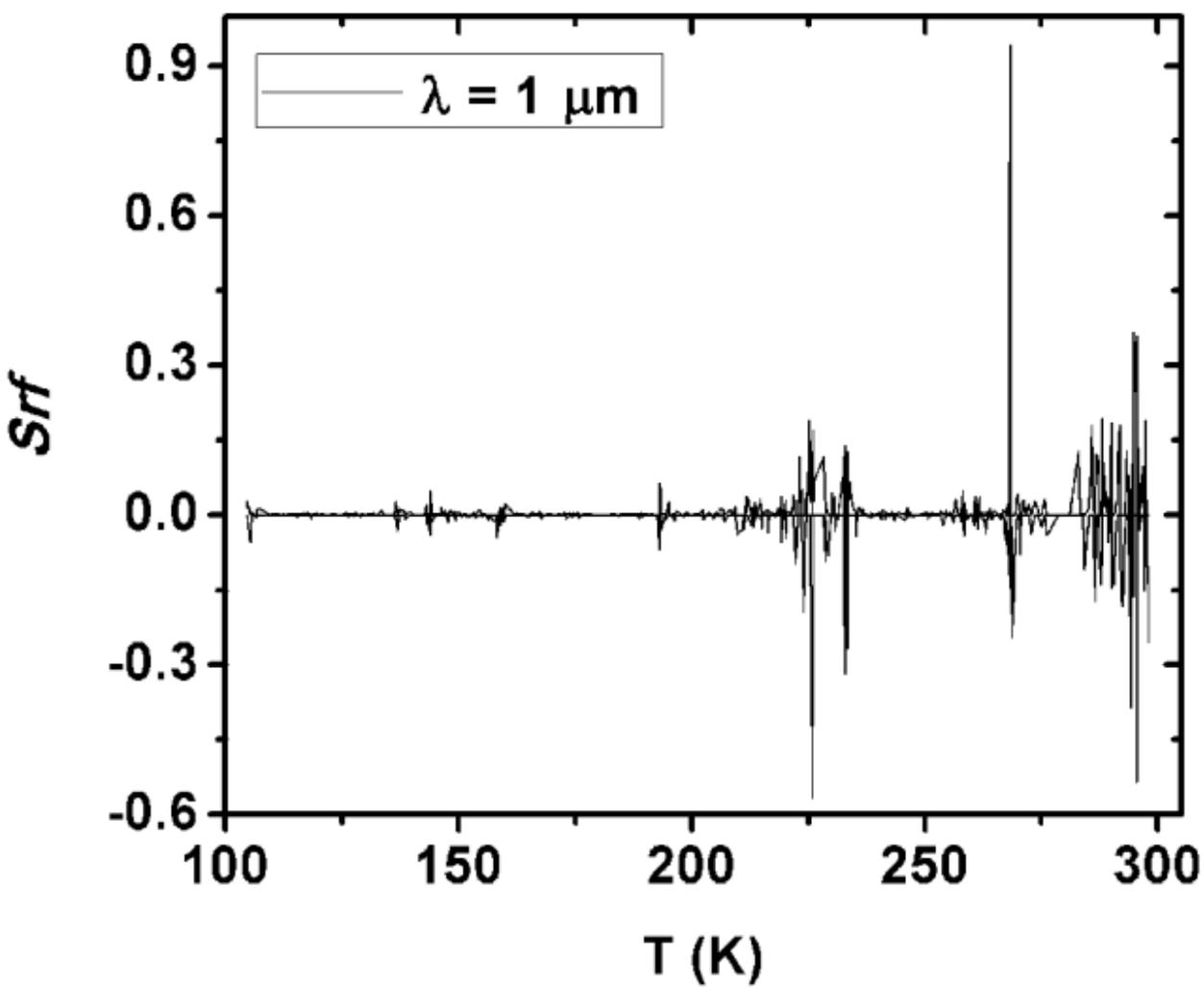

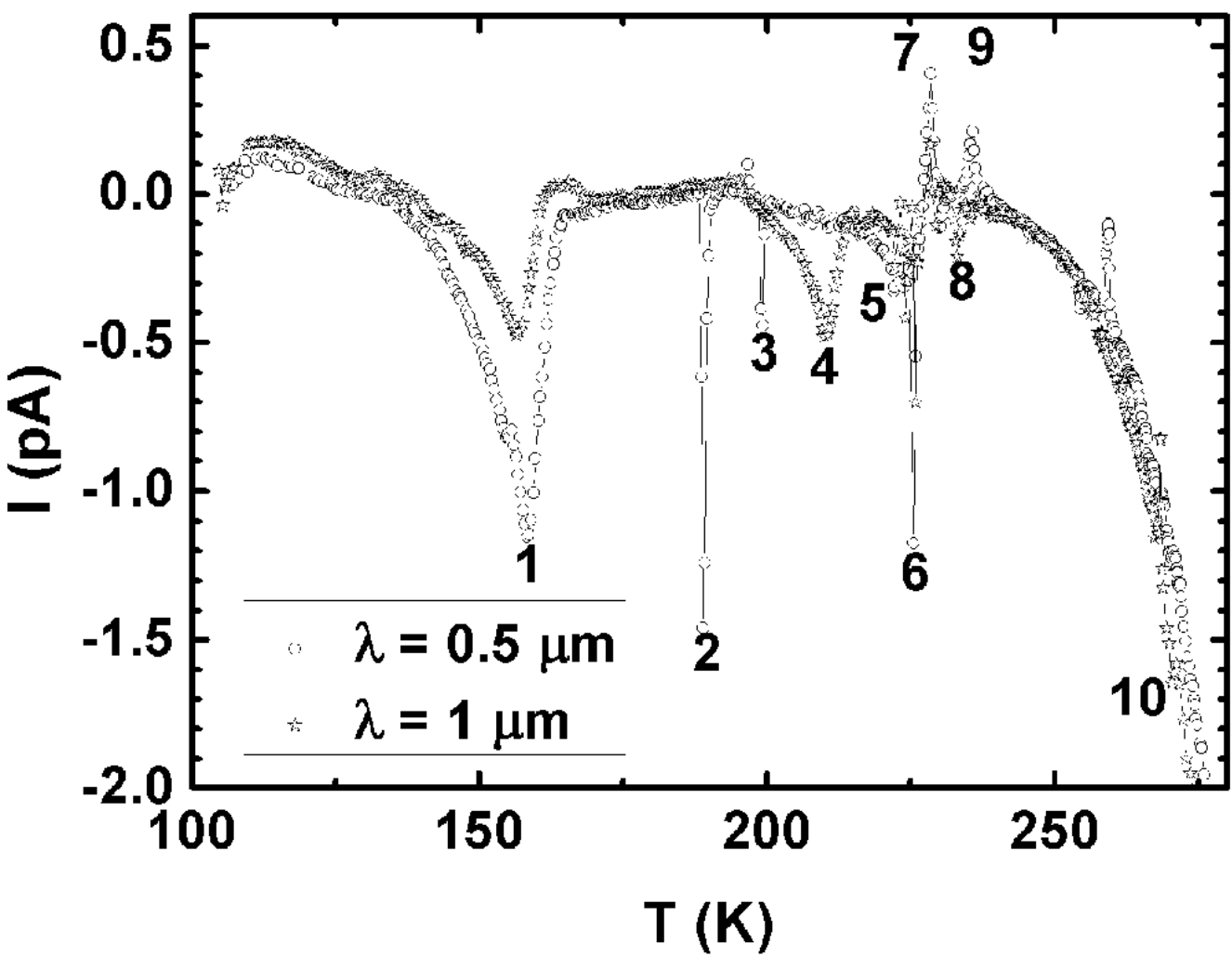

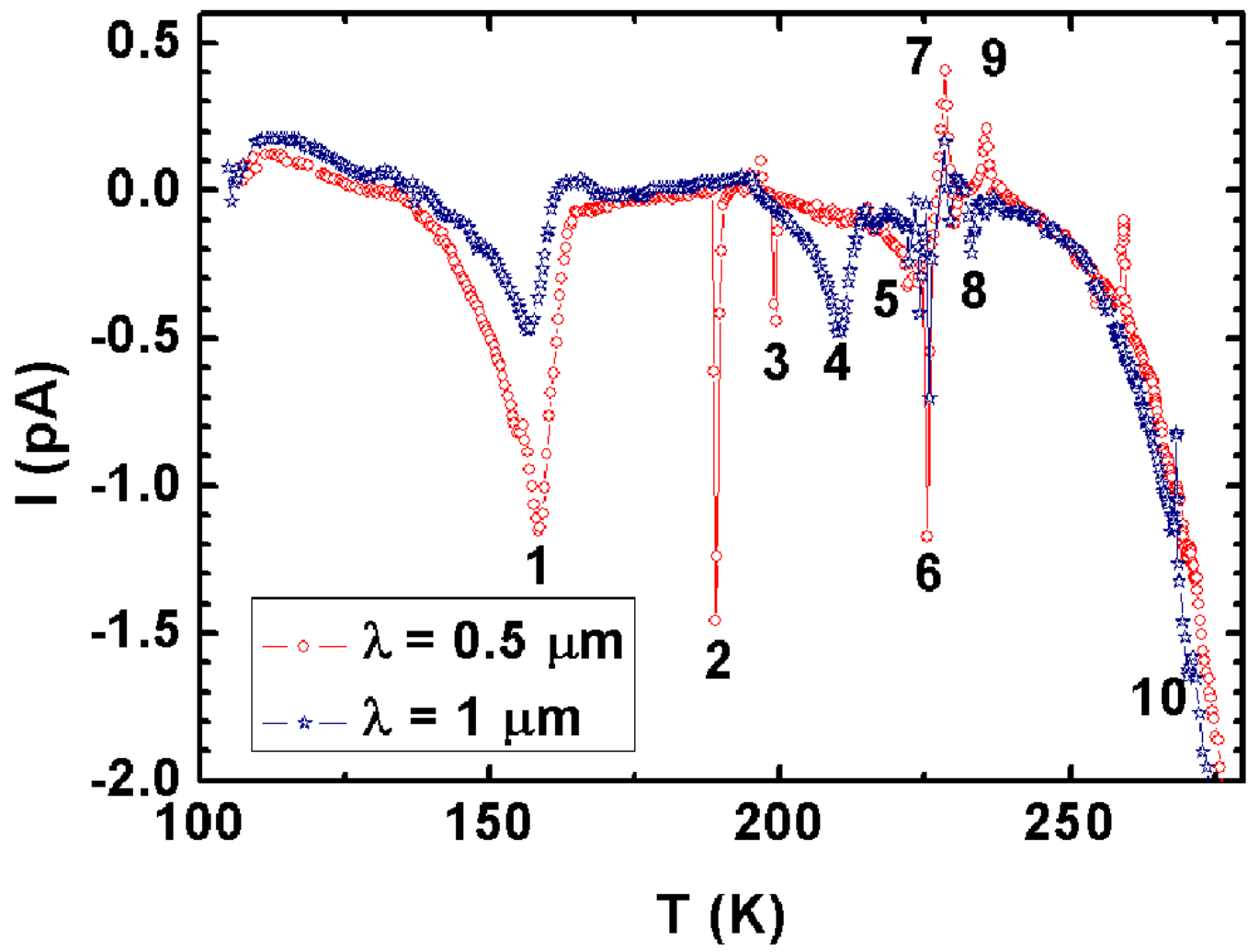

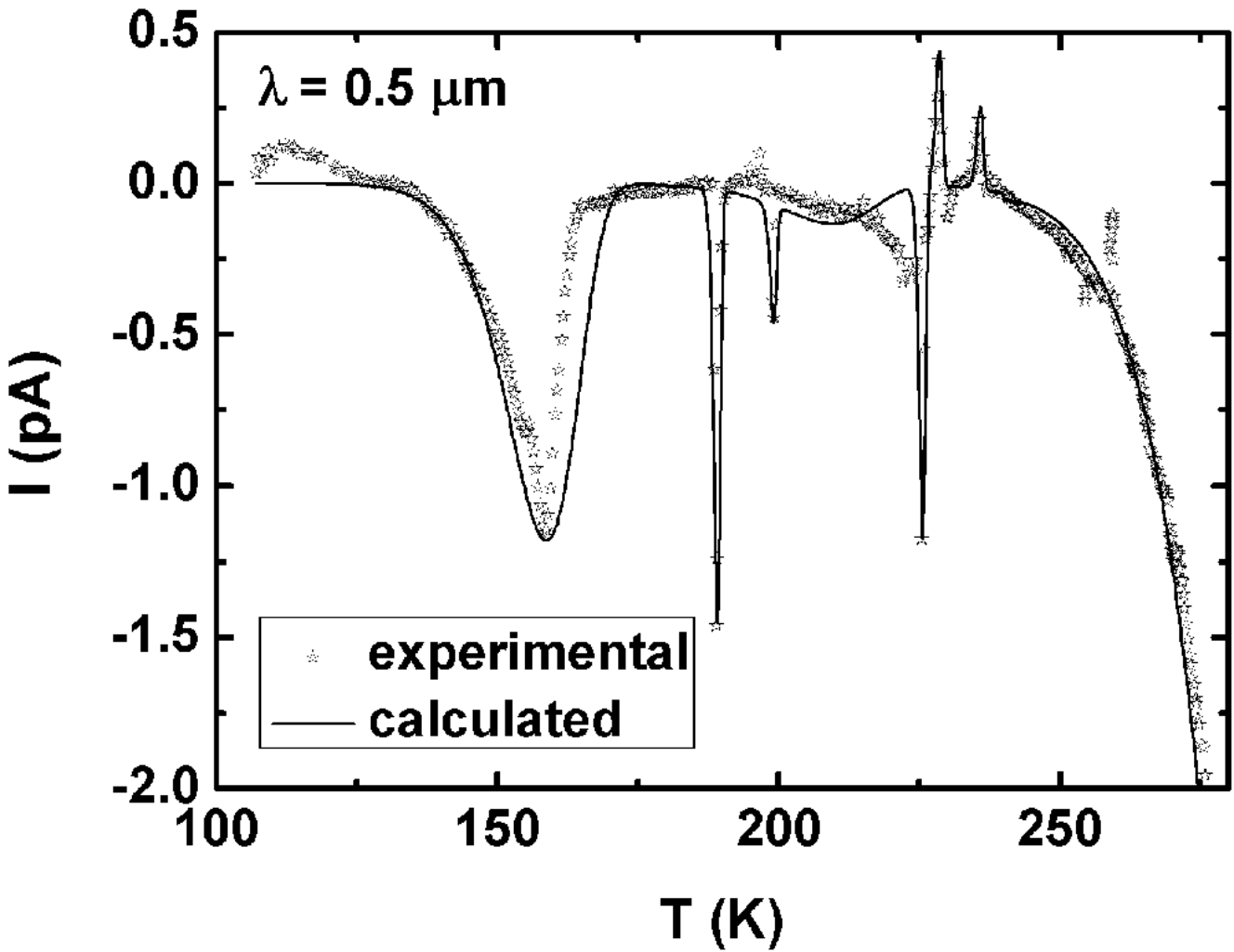

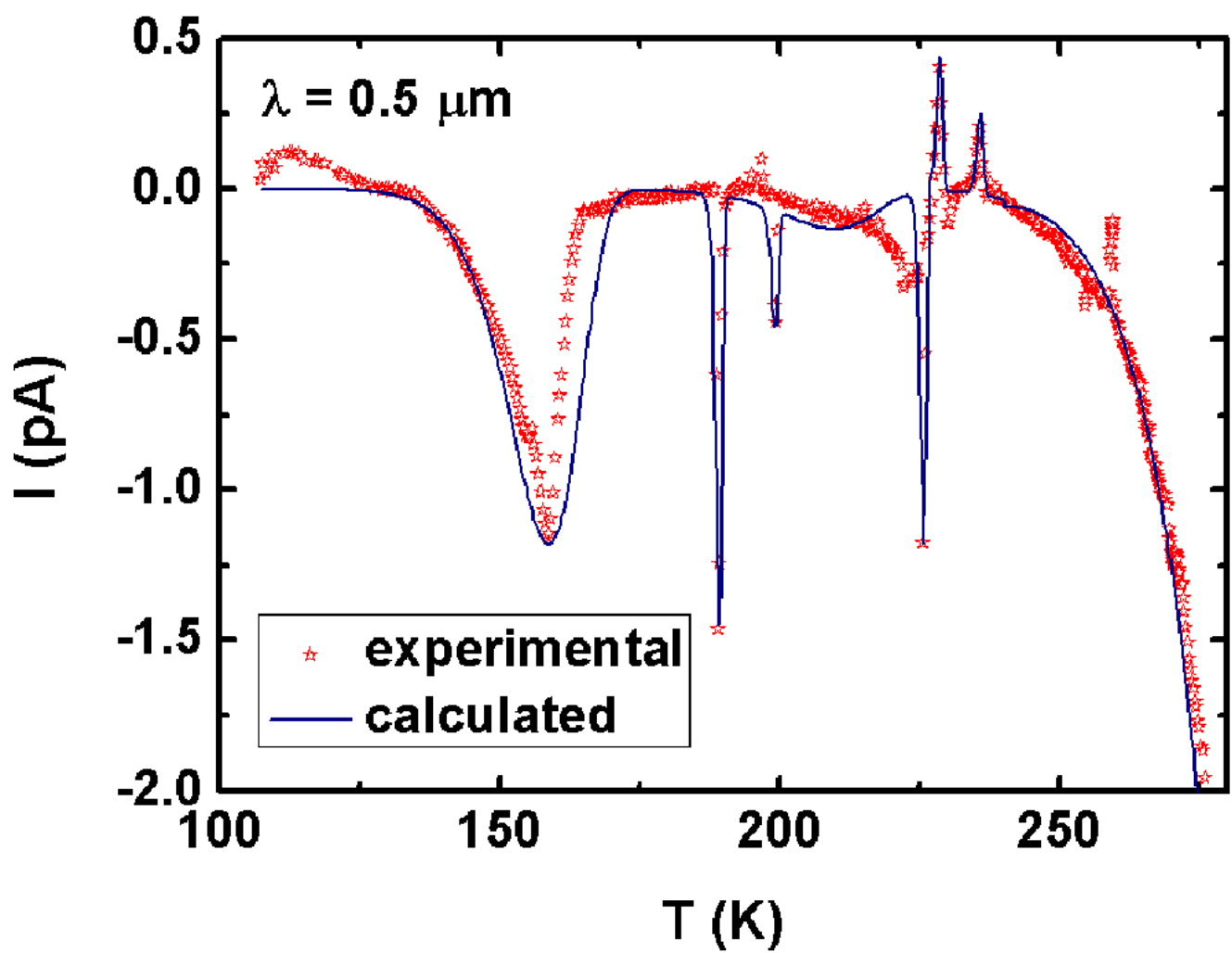

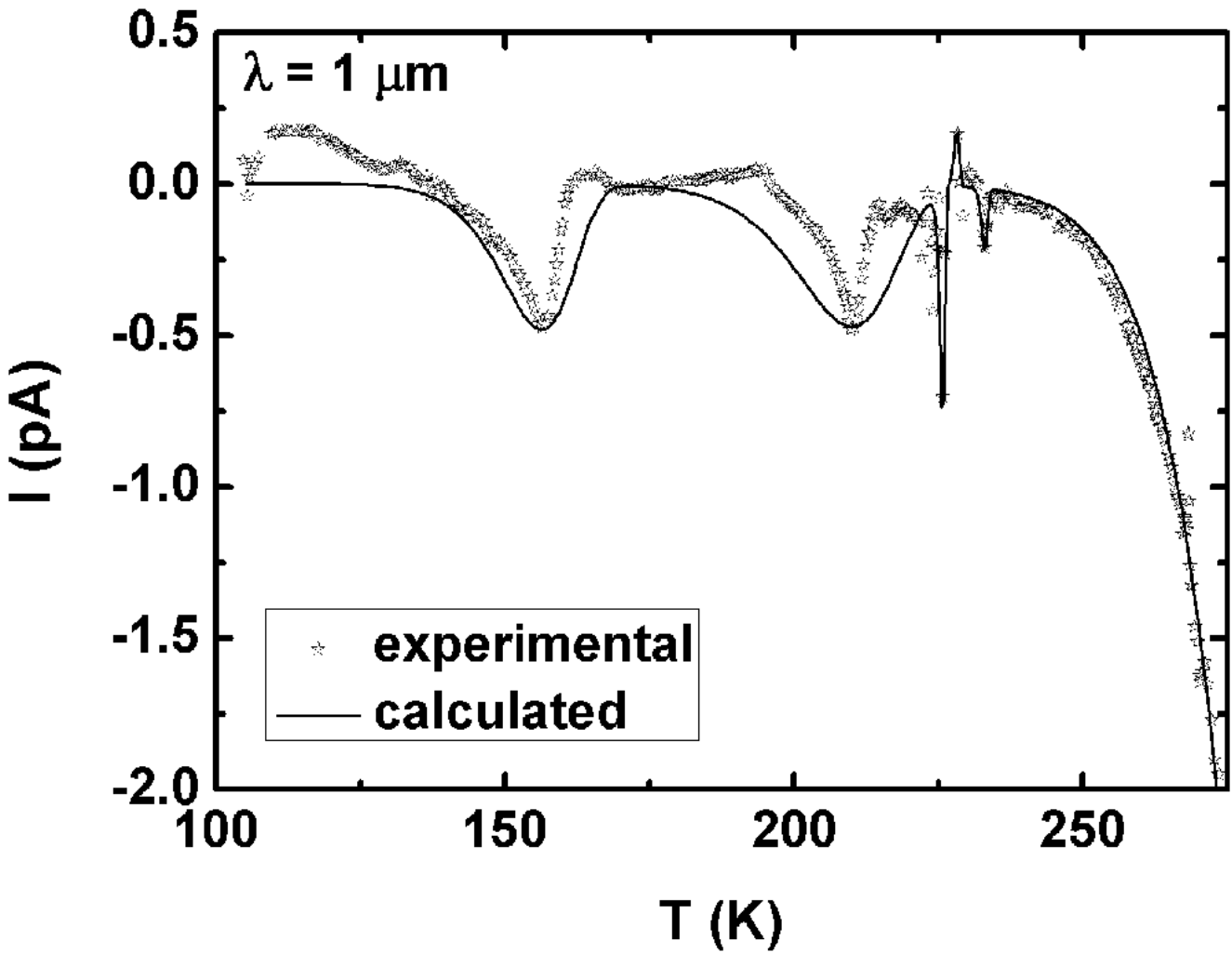

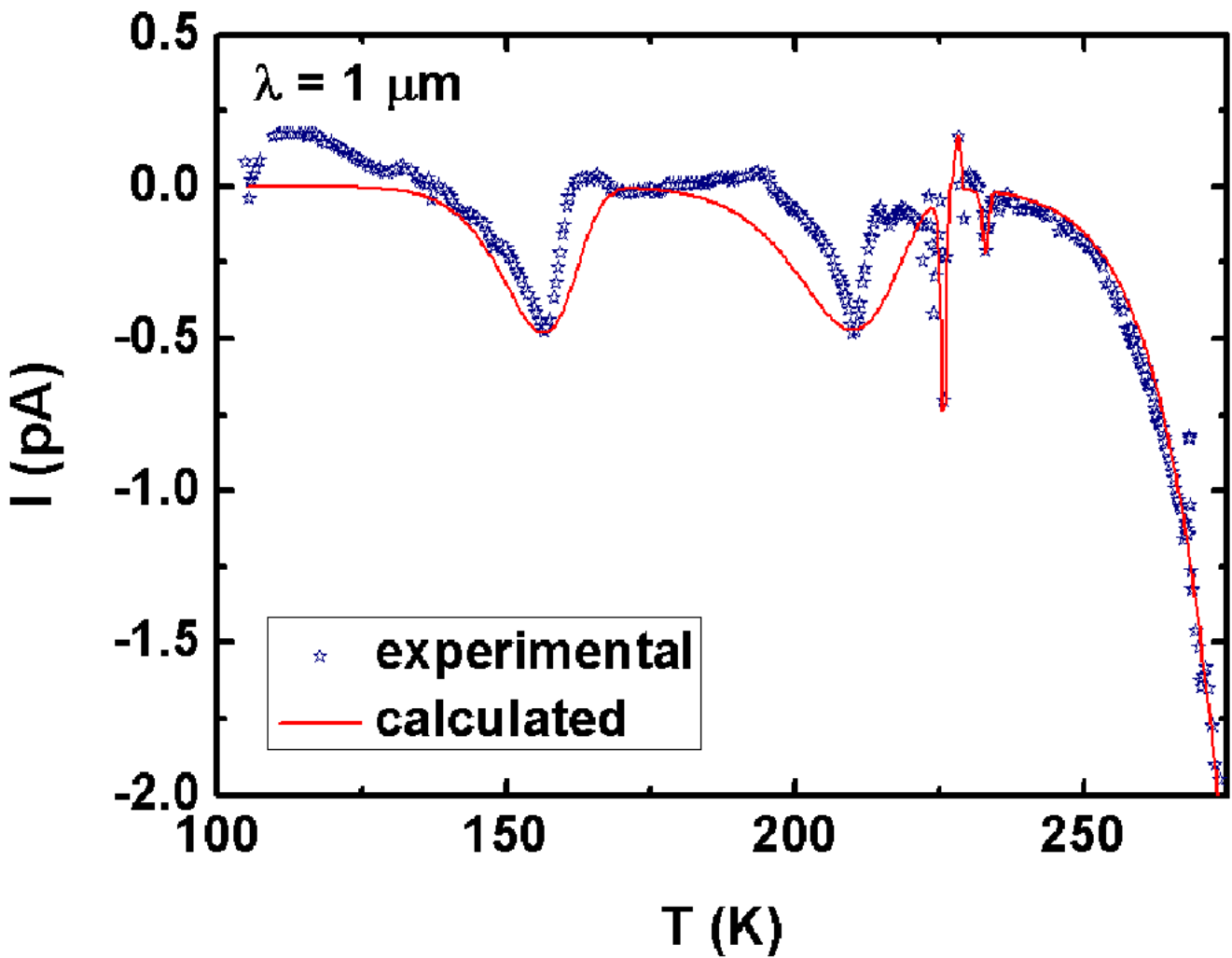